\documentclass[sigconf, nonacm, table]{acmart}

\usepackage[english]{babel}
\usepackage{blindtext}
\usepackage{multirow}
\usepackage{xcolor}
\usepackage{soul}
\usepackage{array}
\usepackage{tagging}
\usepackage{graphicx}
\usepackage{caption}
\usepackage{cleveref}
\usepackage{subcaption}
\usepackage{adjustbox}
\usepackage[most]{tcolorbox}
\usepackage{float}
\usepackage{enumitem}

\usepackage{makecell}    
\usepackage{booktabs}    

\newcommand{\circled}[1]{\tikz[baseline=(myanchor.base)] \node[circle,fill=.,inner sep=1pt] (myanchor) {\color{-.}\bfseries\footnotesize #1};}

\AtBeginDocument{%
  }

\acmSubmissionID{\#17}

\begin{document}

\title{Towards a Playground to Democratize Experimentation and Benchmarking of AI Agents for Network Troubleshooting}



\author{Zhihao Wang}
\affiliation{%
  \institution{UESTC}
  \country{}
}

\author{Alessandro Cornacchia}
\affiliation{%
  \institution{KAUST}
  \country{}
}

\author{Franco Galante}
\affiliation{%
  \institution{Politecnico di Torino}
  \country{}
}

\author{Carlo Centofanti}
\affiliation{%
  \institution{University of L’Aquila}
  \country{}
}

\author{Alessio Sacco}
\affiliation{%
  \institution{Politecnico di Torino}
  \country{}
}
\author{Dingde Jiang}
\affiliation{%
  \institution{UESTC}
  \country{}
}


\renewcommand{\shortauthors}{}


\newif\ifsubmission
\submissionfalse

\ifsubmission
    \newcommand{\ac}[1]{}
    \newcommand{\zw}[1]{}
    \newcommand{\as}[1]{}
    \newcommand{\ms}[1]{}
    \newcommand{\cf}[1]{}
\else
    \newcommand{\ac}[1]{\textit{\color{magenta}[Ales: #1]}}
    \newcommand{\zw}[1]{\textit{\color{cyan}[Zhihao: #1]}}
    \newcommand{\as}[1]{\textit{\color{purple}[Alessio: #1]}}
    \newcommand{\ms}[1]{\textit{\color{red}[Matteo: #1]}}
    \newcommand{\cf}[1]{\textit{\color{green}[Carlo: #1]}}
\fi

\newcommand{\smartparagraph}[1]{\noindent{\bf #1}\ }

\crefname{figure}{Fig.}{Figs.}
\crefname{tabular}{Tab.}{Tabs.}
\crefname{section}{Sec.}{Secs.}
\maketitle

\section{Approach and Motivation}


Recent research has demonstrated the effectiveness of Artificial Intelligence (AI), and more specifically, Large Language Models (LLMs), in supporting network configuration synthesis~\cite{wang2024netconfeval} and automating network diagnosis~\cite{wang2024netassistant} tasks, among others. 
In this preliminary work, we restrict our focus to the application of AI agents to network troubleshooting
and elaborate on \emph{the need for a standardized, reproducible, and open benchmarking platform, where to build and evaluate AI agents with low operational effort}.
This platform primarily aims to \emph{standardize} and \emph{democratize} the experimentation with AI agents, by enabling researchers and practitioners -- including non domain-experts such as ML engineers and data scientists -- to focus on the evaluation of AI agents on curated problem sets, without concern for underlying operational complexities. 
Custom AI agents can be easily plugged through a single Application Programming Interface~(API) and rapidly evaluated. 
We present a modular and extensible benchmarking framework that supports widely adopted network emulators~\cite{bonofiglio2018kathara, mininet, containernet, containerlab}. 
It targets an extensible set of network issues in diverse real-world scenarios -- \emph{e.g.}, data centers, access, WAN, etc. -- and orchestrates the end-to-end evaluation workflows, including failure injection, telemetry instrumentation and collection, and agent performance evaluation.

\smartparagraph{The natural quest for LLMs for network troubleshooting.}
Given a network problem, network engineers need to undertake mechanical yet cumbersome steps to diagnose and mitigate the issue~\cite{wang2024netassistant,netseer}. 
They can be summarized as \emph{(i)} identifying the right telemetry signals to collect, \emph{(ii)} navigating dashboards and interpreting the collected data, 
\emph{(iii)} taking corrective actions based on the insights derived from the telemetry data \emph{(iv)} iterating on the previous speculating about root-causes and \emph{what-if} hypothesis. 
Typically, engineers need to switch on-the-fly across different monitoring configurations, as network issues are ongoing and escalating. Common examples include probing the network to explain packet drops~\cite{netseer, bufscope} or refining the detection logic -- \emph{e.g.}, 
request the collection of more fine-grained queue-length data to zoom into an ongoing incident~\cite{zoom2net}, enable debug-level diagnostics, etc.
This manual process is still complex, slow and error-prone, as it requires expert operators to reason across multiple dimensions.
Furthermore, modern network telemetry paradigms over programmable data planes -- such as sketches~\cite{sketchovsky, dta} and in-band network telemetry (INT)~\cite{int} -- 
have expanded the range of available measurement strategies. The increased expressiveness may introduce additional degrees of freedom and come at the cost of greater operational complexity.
Thus, we observe that while programmable data planes and new telemetry techniques have enhanced operators' visibility on the network, human intervention still remains a primary bottleneck in network triaging. 

Thanks to their ability to parse multimodal data and, more recently, engage in natural-language-driven reasoning~\cite{wang2024netconfeval, wang2024netassistant, netllmbench}, LLMs -- as a breakthrough category of AI models -- hold a special promise for assisting network operators. 
As a result, our community has begun to explore how traditional approaches can evolve into more automated, LLM-assisted \emph{intent-based} network monitoring and diagnosis solutions~\cite{angi2025llnet, habib2025llm, li2025llmsketch, moghadas2025stradallm}.


\smartparagraph{Lack of holistic platforms and benchmarks.}
Nevertheless, existing experimentation environments~\cite{bonofiglio2018kathara, containerlab} are often limited in scope, lacking standardized and reproducible benchmarks.
Recently, benchmarks for LLM agents have been proposed in the area of network configuration, such as NetConfEval~\cite{wang2024netconfeval}. 
This work assumes that LLM agents can be evaluated on static benchmarks, with one-shot and offline executions. 
While this assumption holds true for many problem instances in network configuration, \emph{network troubleshooting} is an inherently more dynamic and interactive problem space. It requires real-time feedback loops with the network, where AI agents must not only observe but also probe, react, and refine based on the evolving system conditions. 
Thus, it is essential to evaluate AI agents in environments that allow for such interactive, closed-loop operations, where they can dynamically adapt their strategies based on real-time telemetry and network state.
Unfortunately, engineers wishing to experiment with AI agents for network troubleshooting must still shoulder the significant burden of developing custom infrastructure code to enable this agent-environment interaction, as no modular, low-effort platform -- abstracting network environments (e.g., cloud, edge, mobile, satellite), protocols, failure modes, and tooling -- exists to date.

We draw attention to the risk that, in the absence of standardized evaluation practices, research in this area remains fragmented and difficult to compare.
In response, we argue for a holistic, universally accepted platform that combines an extensible benchmark suite and a flexible, plug-and-play network playground -- e.g., based on existing network emulators. We develop a preliminary Proof-of-Concept~(PoC) and validate it with a ReAct agent for a toy-case failure scenario. We then outline challenges and future directions towards our vision.

\begin{figure*}[!t]
    \centering
    \includegraphics[width=\linewidth]{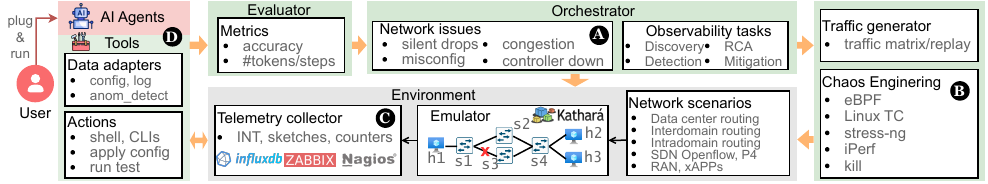}
    \caption{Architecture of the proposed framework.}
    \label{fig:architecture}
\end{figure*}

\section{Proof-of-Concept}

\begin{figure}[!t]
\centering
\begin{tcolorbox}[colback=blue!5, 
colframe=blue!75!black,
fontupper=\scriptsize,
width=\linewidth,
boxrule=0.4pt]

\textsc{\textbf{\textcolor{black!70!black}{User}}:}  
You are an expert networking engineer who has been tasked with detecting anomalies in a deployed network topology: s1:{'name': 'eth0', 'port': 1, 'connected\_to': 'h1', 'connected\_port': 1}[\ldots].You will begin by analyzing the network's state, and then submit your solution:[\ldots] You are provided with the following APIs to interact with the network: [..]

\textsc{\textbf{\textcolor{red!70!black}{ReAct Agent}}:}
\textit{\textbf{Thought}:} 
To begin anomaly detection, I'll first check the overall network reachability [\ldots]
\textit{\textcolor{green!70!black}{\textbf{Action}:}} 
\texttt{test\_reachability()}

\textsc{\textbf{\textcolor{blue!70!black}{Framework}}:}  
`h1 ping h2: 10 packets transmitted, 10 received, 0\% packet loss', `h1 ping h3: 10 packets transmitted, 0 received, 100\% packet loss' [\ldots]

\textsc{\textbf{\textcolor{red!70!black}{ReAct Agent}}:}
\emph{\textbf{Thought:}} 
[\ldots] the problem may lie in the s1-s3 link or the s3-s4 link. I'll start by checking port counters on s1's port connected to s3 (port 3) to detect potential packet loss. 
\textit{\textcolor{green!70!black}{\textbf{Action:}}}  \texttt{bmv2\_get\_counters("s1")}

\textsc{\textbf{\textcolor{blue!70!black}{Framework}}:}  
`MyIngress.ingress\_port\_counter', `MyEgress.egress\_port\_counter'

\textsc{\textbf{\textcolor{red!70!black}{ReAct Agent}}:}
\textit{\textbf{Thought}:} 
[\ldots] Since we're investigating the path from h1 to h3, which should go through s1's port 3 (egress to s3), I'll check the egress counter for port 3 to see if packets are being sent out properly.
\textit{\textcolor{green!70!black}{\textbf{Action}:}} 
\texttt{bmv2\_counter\_read("s1", "MyEgress.egress\_port\_counter", 3)}

\textsc{\textbf{\textcolor{blue!70!black}{Framework}}:}  
MyIngress.ingress\_port\_counter[3]= (980 bytes, 10 packets)

{
\smallskip
\centering 
\emph{The agent checks more counters} 
\smallskip
\par
}

\textsc{\textbf{\textcolor{red!70!black}{ReAct Agent}}:}
\textit{\textbf{Thought}:} 
[\ldots] suggest a potential unidirectional link failure or port malfunction specifically affecting traffic from s3 to s1.[\ldots]
\textit{\textcolor{green!70!black}{\textbf{Action}:}} 
\texttt{submit\_findings()}
\end{tcolorbox}

\resizebox{\linewidth}{!}{%
\footnotesize
\begin{tabular}{c c l}
    \textbf{Category} & \textbf{API} & \textbf{Description} \\
    \midrule
    \multirow{5}{*}{\makecell[c]{Data\\Adapters}} 
    & \texttt{get\_switch\_logs/info} & Get device running logs/information  \\
    & \texttt{ovs/bmv2\_dump\_ports} & Show all ports of OVS/Bmv2 P4 switch\\
    & \texttt{bmv2\_get\_counters} & Get counters in a BMv2 P4 switch \\
    & \texttt{bmv2\_counter\_read} & Read counter values in a BMv2 P4 switch \\
    & \texttt{get\_topology} & Obtain structured topology information\\
    \midrule

    \multirow{4}{*}{Actions} 
    & \texttt{config\_frr\_bgp/ospf} & Configure BGP/OSPF in FRRouting \\
    & \texttt{ovs\_table\_add/modify} & Add/modify flow table entry of OVS \\ 
    & \texttt{bmv2\_table\_add/modify} & Add/modify table entry in BMv2 P4 switch \\
    & \texttt{test\_reachability} & Check reachability between all hosts \\
\end{tabular}
}

\caption{Tools implemented in our PoC and agent trajectory.}
\label{fig:agent}
\end{figure}

An overview of our playground's architecture is shown in Figure~\ref{fig:architecture}.
Based on this design, we prototype an initial PoC\footnote{codebase will be open-sourced} on top of Kathara~\cite{bonofiglio2018kathara} and validate an end-to-end network failure scenario, which we triage using a ReAct~\cite{yao2023react} AI agent.
The user (e.g., ML engineer) selects one of the pre-defined network issues \circled{A} through a declarative interface.
The user can implement the AI agent logic simply by rewriting a callback \texttt{execute\_agent}. Then, the user can plug the code into our platform and kickoff the evaluation workflow in a single command. The playground is responsible for instantiating the experiment and orchestrating the underlying operations such as traffic generation, fault (or misconfiguration) injection \circled{B}, telemetry instrumentation and telemetry collection \circled{C}. 
The agent can then interact with the network environment through a set of MCP-based~\cite{mcp2025} tools \circled{D} 
that expand its capabilities. Our design follows related work~\cite{aiopslab} in microservice applications.

In our toy-case example, a \texttt{DeepSeek-R1-0528} agent tries to triage a network issue, whose root cause is a lossy link.
We inject an artificial packet loss issue on the $\textsf{s1} \rightarrow \textsf{s3}$ link in the topology shown in the emulator box of \cref{fig:architecture}, which comprises four BMv2 switches.
We task the agent with (1) detecting and (2) localizing the anomaly. 
\cref{fig:agent} (top) illustrates the agent's reasoning trajectory, overall consisting of 15 steps.
The agent is prompted with the operator's intent (\textsc{User} line) and is given no clue about the anomaly root-cause. It can access the tools illustrated in \cref{fig:agent} (bottom) to interface with the environment.
It begins with active probing via \texttt{get\_reachability()}, detects loss between \textsf{h1} and \textsf{h3}, then proceeds to querying port counters using \texttt{bmv2\_counter\_read()}. Based on the retrieved statistics, it successfully localizes the fault to \textsf{s3}.

\section{Future Agenda}

\smartparagraph{Benchmark curation.}
We aim to curate a diverse benchmark of failure scenarios, spanning heterogeneous networks (e.g., data centers vs. geographical networks) network stacks and failure type. Each scenario is manually constructed with defined triggers, observability signals (e.g., INT latency spikes, counter anomalies) and root causes. 
A primary challenge in curating such a diverse benchmark lies in minimizing human effort while ensuring sufficient scenario \emph{variations}. We plan to study how to automate the generation of these variations, starting from a well-defined subset of network issues across different domains. To this end, we can draw inspiration from similar approaches in software engineering~\cite{jain2024re}.
Furthermore, we plan to explore automatic ways of tuning the level of complexity of the injected problem sets, e.g., by tweaking temporal patterns or combining multiple failures. It can be accomplished via parametric failure injection templates~\cite{chaos-mesh}, or, for other class of failure modes, we could explore LLMs themselves to generate failure modes by reasoning on configuration files and network setups. 

\smartparagraph{Agent–environment interfaces.}
We envision developing unified agent–environment interfaces that abstract low-level complexity and expose structured access to both telemetry (e.g., system metrics, INT, sketches) and control (e.g., configuration updates, active probing). We aim to align these interfaces with MCP to support structured context exchange and standardized agent–environment interaction.
Prior work has shown that equipping AI agents with task-specific tools is critical for diagnostics like RCA~\cite{wang2024rcagent}. Rather than parsing raw, heterogeneous telemetry directly with AI agents, agents invoke and interpret outputs from modular tooling, such as ML-based anomaly detectors. This design mitigates the limitations of AI agents in handling domain-specific tasks and enables more robust, composable reasoning pipelines.


\smartparagraph{Automated assessment of agent behavior.}
The analysis results generated by LLMs are generally presented as unstructured natural language descriptions, often requiring operators to manually assess their accuracy.
The manual inspection of an agent's execution trajectory (\cref{fig:agent}) is time-consuming, limits experiments scalability, and obstructs fairness and reproducibility. To address the gap, we propose extending our framework with automated behavioral checkups -- e.g., leveraging \emph{LLM-as-a-judge}~\cite{li2025generationjudgmentopportunitieschallenges} -- to evaluate agent trajectories in a more structured and holistic manner. 
As new tools for observability of agentic AI start emerging~\cite{langfuse, langsmith}, providing richer logging, monitoring, and performance metrics, we plan to build upon these tools to systematically trace, record, and debug LLM-agent executions and support automated trajectory analysis.
This opens the door to more meaningful diagnostics and supports downstream uses, such as targeted fine-tuning.


\bibliographystyle{ACM-Reference-Format}
\bibliography{references}  

\end{document}